\documentclass[11pt]{article}
\usepackage{amsmath,amsthm,latexsym,amssymb,amsfonts,epsfig}
\usepackage{fontenc,dsfont,layout}
\numberwithin{equation}{section}
\newcommand{\Na}{\mathbb{N}}      
\newcommand{\Do}{\mathcal{D}}               %domain$
\newcommand{\C}{\mathbb{C}}               % the complex numbers
\def\H{{\cal H}}

\DeclareMathOperator{\porder}{\mathcal{P}}             % pathorder operator
\newcommand{\bra}[1]{|{\bf{#1}}\rangle}                        % bracket
\newcommand{\ket}[1]{\langle{\bf{#1}}}
\newcommand{\one}{\text{\bf 1}}               % identity operator

\def\bazaa{{}^o\! e}
\def\bazada{{}^o \! \omega}
\newcommand{\baza}[2]{^o \! e^{#1}_{#2}}
\newcommand{\bazad}[2]{^o \! \omega^{#1}_{#2}}

\newcommand{\M}{\Sigma}
\def\rig{{}^o\!\eta}
\def\q{{}^o\!q}

\def\SU(2){{\rm SU(2)}}

\newcommand{\be}{\begin{equation}}
\newcommand{\ee}{\end{equation}}

\begin{document}

\title{An open FRW model in Loop Quantum Cosmology}

\author{{\L}ukasz Szulc \thanks{lszulc@fuw.edu.pl} 
\date{\it  Institute of Theoretical Physics, Warsaw University
ul. Ho\.{z}a 69, 00-681 Warszawa, Poland \\
[.5cm]}} \maketitle

\begin{abstract}
Open FRW model in Loop Quantum Cosmology is under consideration. The
left and right invariant vector fields and holonomies 
along them are studied. It is shown that in the hyperbolic geometry of $k=-1$ itis possible to construct a suitable loop which provides us with quantum
scalar constraint originally introduced by Vandersloot \cite{17}.
The quantum scalar constraint operator with negative cosmological constant is proved to be essentially self-adjoint. 
\end{abstract}

\section{Introduction}
Loop Quantum Cosmology (LQC) is a novel approach to quantum theory of
cosmology \cite{4,5,6}. The Loop Quantum Gravity (LQG) \cite{1,2,3} inspired quantization of the symmetry reduced models which are the test field for the full theory. It also provides some very interesting results like quantum geometry effects and absence of singularity \cite{10}.
During last years there has been a progress in the area of
LQC \cite{9,10,12,13,26}. Although the part of isotropic ($k=0$
\cite{15} and $k=1$ \cite{9,18,19}) and
homogeneous sector of quantum theory is well understood there still is a
problem with Bianchi class B models such as open Friedmann
Robertson Walker (FRW) model (so called $k=-1$)\footnote{Strictly
speaking the isotropic $k=-1$ model is derived from anisotropic
Bianchi V}. One of the reasons for that is the well known
problem concerning the Hamiltonian formulation of class B models. However,
one can derive that the isotropic $k=-1$ model in terms of real Ashtekar
variables has correct Hamiltonian formulation (see \cite{17} and references therein). The second problem
comes form the geometric difficulty. It is not clear how to
introduce a loop suitable for the quantization purposes. 
Although there has been a recent progress in the FRW hyperbolic model
\cite{17} a potential gap arose. The object which was quantized in \cite{17} wasthe holonomy along an open curve. However, such a holonomy should not be considered as the
components of the curvature. Moreover, the holonomies considered in
\cite{17} were used with respect to the $\gamma K=A-\Gamma$ variables rather
then to the $A$ connection (there is nothing wrong with it logically, but it
makes the relation to the full theory obscure). However, quantum theory described in
\cite{17} does not suffer from singularities when gravity is coupled to
homogeneous massless scalar field and has a correct classical limit as well. In that sense the theory of quantum $k=-1$ model is correct and provides new quantum gravitational effects. One can conclude that
\cite{17} is a major candidate to replace the old classical $k=-1$ model by the
new quantum one. We show in the present paper how this model can become
conceptually closer to the full (LQG) theory by intoducing a suitable loop, which leads to the same scalar constraint as in \cite{17}. Using a new loop in quantum theory has one more advantage. The implementation of ``improved dynamics'' introduced by Ashtekar, Paw{\l}owski and Singh in \cite{15} is direct and natural what is missing in \cite{17}. However, our results are valid
again only for the $\gamma K$ holonomies.

%one of the main problems was not resolved. There was
%simply no loop to quantize. The object which was quantized in \cite{17} was
%the holonomy along an open curve which should not be considered as components of the curvature. Such a theory has several disadvantages such as lack
%of gauge invariance of the trace of holonomy along an open curve.
%However is was shown, that quantum theory in
%\cite{17} does not suffer from singularities when gravity is coupled to
%homogeneous massless scalar field. Such a quantum theory has a correct
%semiclassical limit. Natural question then arises: what is the
%geometrical interpretation of the corresponding $k=-1$ quantum theory? How should
%we think about the quantum evolution if quantum Hamiltonian is not
%ruled by quantum loop? 

%All these questions might seem troublesome for the theory, however we will show
%in this paper, that this is not the case. The quantum scalar
%constraint operator introduced by Vandersloot in \cite{17} has indeed
%correct geometrical interpretation (is constructed from the suitable
%loop). Moreover, the physical area of the new loop can be constrained to
%be minimal as in previous isotropic models
%\cite{15,18,19}. Collecting results from \cite{15,17,18,19}, as well as from
%this paper, one can conclude that suitable geometric quantization of
%three isotropic and homogeneous models $k=0,\pm 1$ leads to the common
%kinematical (and in the case of minimally coupled massless scalar
%field also physical) Hilbert space.

This paper is organized as follows. In section 2 we briefly discuss
the classical hyperbolic geometry of $k=-1$ model as well as its
Hamiltonian formulation. In section 3 the quantum
scalar constraint is constructed in detail by introduction of a suitable
loop. Section 4 describes properties of the scalar constraint operator with
negative cosmological constant.

\section{Classical Theory}
\subsection{FRW models}
The well known isotropic and homogeneous sector of the General Relativity can be
considered as three so called Friedmann-Robertson-Walker models, where the
metric tensor has a form
\begin{equation}\label{FRW}
g= -N^2(t)dt^2 + a(t)^2 [(1-kr)^{-1} dr^2 + r^2d\Omega^2].
\end{equation}
Because of the large number of symmetries there is only one gravitational degree of freedom, the scale factor $a(t)$ which is a function of arbitrarily chosen time coordinate $t$. $N(t)$ is referred to as a laps function and does not enter the equations of motion as a dynamical variable. $k$ is a number which can take only three values. Each of the values of $k$ corresponds to different intrinsic curvature of spatial manifold $\Sigma$. Spatially flat, closed and open universes are described by $k=0,+1, -1$ respectively. Einstein equations for (\ref{FRW}) describe the dynamics of the scale factor
\begin{equation}\label{Friedmann}
\left( \frac{\dot{a}}{a} \right)^2 + \frac{k}{a^2} = \frac{8 \pi G}{3} \rho_{\rm matter}.
\end{equation}
This equation describes evolution of the universe filled by matter density $\rho_{\rm matter}$. 

\subsection{The $k=-1$ geometry}
It is well known \cite{27} that spatial part of (\ref{FRW}) can be written in terms of left invariant one-forms as
\begin{equation}\label{qmetric}
q=a^2(t) \delta_{ij} \ \bazad{i}{a} \ \bazad{j}{b} \ dx^a dx^b
:=a^2(t) \q_{ab} dx^a dx^b ,
\end{equation}
where $i,j=1,2,3$. These one-forms $\bazad{i}{a}$ satisfy Maurer-Cartan equation
\begin{equation} \label{MC}
\partial_a \ \bazad{i}{b} = -\frac{1}{2}  {C^i}_{jk} \ \bazad{j}{a} \ \bazad{k}{b} ,
\end{equation}
where for $k=-1$ structure constants are given by 
\begin{equation}\label{structure} 
{C^k}_{ij}=\delta^k_i \delta_{j1} - \delta^k_j \delta_{i1} .
\end{equation}
The same structure constants occur in the algebra of left invariant vector fields \footnote{Left invariant one-forms are dual to left invariant vector fields: $ \bazada^i (\bazaa_j)=\delta^i_j $, where $\bazada^i=\bazada^i_a dx^a$ and $\bazaa_i=\bazaa^a_i \partial_a$} on $\Sigma$ ( it is well known Bianchi class V algebra) 
\begin{equation}\label{algebra}
[ \bazaa_i , \bazaa_j ] = {C^k}_{ij} \bazaa_k .
\end{equation}
Left invariant one-forms and vector fields can be written in some coordinates $x^a$ as
\begin{align}\label{exact}
\bazaa_1 &= \partial_1 \quad \quad \quad \quad \bazada^1=dx^1 ,\nonumber \\
\bazaa_2 &= e^{-x^1} \partial_2 \quad \quad \bazada^2 = e^{x^1} dx^2 ,\\
\bazaa_3 &= e^{-x^1} \partial_3 \quad \quad \bazada^3 = e^{x^1} dx^3 .\nonumber
\end{align}
One can check that equations (\ref{algebra}) and (\ref{MC}) are
satisfied by (\ref{exact}). Note, that ${C^i}_{ij}\neq 0 $. Such algebras belong to the class B in the Bianchi classification.

%It is well known that Bianchi class B models do not have a
%Hamiltonian formulation consistent with the Einstein equations (in general
%Hamiltonian equations of motion differ from Einstein
%equations). However, as is pointed in \cite{17} the isotropic Bianchi V model
%in terms of Ashtekar variables has correct Hamiltonian formulation.

\subsection{Classical Dynamics}
Canonical quantization of full General Relativity as well as symmetry
reduced models is based on their Hamiltonian formulation \cite{1,3}. In terms of Ashtekar variables the full
Hamiltonian for GR is a sum of constraints
\begin{equation}
H_{\rm gr}^{\rm tot}= \int_{\Sigma}d^3x (N^i G_i + N^a C_a + N  h_{\rm sc}),
\end{equation}
where 
%C_a &= E^b_i F^i_{ab} - (1-\gamma^2)K^i_a G_i ,\nonumber \\
\begin{align}
C_a &= E^b_i F^i_{ab}, \nonumber \\
G_i &= D_a E^a_i := \partial_a E^a_i + {\varepsilon_{ij}}^k A^j_a E^a_k
\end{align}
are called diffeomorphism and Gauss constraints respectively. $F=dA + \frac{1}{2}[A,A]$ is a
curvature of Ashtekar connection (\ref{variables}). The most complicated
scalar constraint has a form
\begin{align}\label{ham}
&H_{\rm gr}:=\int_{\Sigma}d^3x N(x) h_{\rm sc}= \nonumber \\ &\frac{1}{16 \pi G} \int_{\Sigma} d^3x N(x)\left( \frac{E^a_i
  E^b_j}{\sqrt{|\mathrm{det} E|}} {\varepsilon^{ij}}_k F_{ab}^k -
  2(1+\gamma^2) \frac{E^a_i E^b_j}{\sqrt{|\mathrm{det} E|}} K^i_a
  K^j_b \right) .
\end{align}
The Ashtekar variables $(A,E)$ are constructed from the triads (see
\cite{1,3} for details) in the following way
\begin{equation}\label{variables}
A^i_a = \Gamma ^i_a + \gamma K^i_a \quad \quad E^a_i =
\sqrt{|\mathrm{det}q|}e^a_i ,
\end{equation}
where $q_{ab}=\delta_{ij} e^i_a e^j_b$, and $\mathrm{det}q$ stands for the 
determinant of spatial metric $q_{ab}$. $A$ and $E$ take values in
su(2) algebra and $\mathrm{su(2)}^{*}$ dual algebra respectively.
In the case of symmetry reduced model (for the case $k=-1$ see (\ref{qmetric})) the above
equations simplify dramatically. (\ref{variables}) reduce to 
\begin{equation}\label{reduced}
A^i_a =- {\varepsilon^{1i}}_j \ \bazad{j}{a} + \tilde{c} \ \bazad{i}{a} \quad \quad E^a_i= \tilde{p}
\sqrt{\mathrm{det}\q} \ \baza{a}{i}  ,
\end{equation}
where $\tilde{c}=\gamma \dot{a}$ and $\tilde{p}=a^2$. Notice that connection $A$ is
not diagonal. This is very different situation than $k=0$ 
and $k=1$. One can check, that Gauss and diffeomorphism constraints in variables (\ref{reduced}) 
are satisfied automatically. The only non-trivial one is scalar constraint. From (\ref{ham}) and (\ref{reduced}) we get 
\begin{equation}
H_{\rm gr}=-\frac{3 V_0}{8 \pi G \gamma^2} \sqrt{|\tilde{p}|} (\tilde{c}^2 -
\gamma^2) ,
\end{equation}
where $\tilde{c}$ and $\tilde{p}$ are canonically conjugated $\{\tilde{c},\tilde{p}\}=\frac{8\pi
  G \gamma}{3V_0}$ and $N(t)=1$. $V_0$ is a volume of elementary cell (see
  \cite{15,19,18} for details).
In the presence of matter (in the isotropic and homogeneous case) the
  term $H_{\rm matt}=V_0 |\tilde{p}|^{3/2} \rho_{\rm matt}$ is added to gravitational scalar constraint
\begin{equation}
H^{\rm tot}=-\frac{3 V_0}{8 \pi G \gamma^2} \sqrt{|\tilde{p}|} (\tilde{c}^2 - \gamma^2)+ V_0 
|\tilde{p}|^{3/2} \rho_{\rm matt}.
\end{equation}
If $H^{\rm tot}$ is constrained to vanish, one can easily check that the Friedmann
equation (\ref{Friedmann}) is recovered (for $k=-1$). We showed then
followed by \cite{17} that indeed the isotropic
Bianchi V (class B) model has correct Hamiltonian formulation.

\section{Quantum Theory}
\subsection{Kinematics}
Quantum kinematics in the full Loop Quantum Gravity is based on
the classical Poisson bracket algebra between holonomy along an edge
$h_e[A]$ and fluxes ($E$ smeared on 2-surface) $E(S)$ \cite{3,1}. In
the isotropic and homogeneous models $k=0,1$ holonomies are
reduced to the so-called almost periodic functions $\sum_{\mu} \xi_{\mu}
e^{\frac{i\mu c}{2}}$.  Classical algebra $\{\tilde{p},e^{\frac{i\mu \tilde{c}}{2}}\}$
is then easy to quantize and quantum theory is placed in the Bohr
compatification of a real line \cite{12,13,19,18}. In the case of $k=-1$ the
situation is more complicated \cite{17}. Because the $A$ connection
is no longer diagonal the classical Poisson bracket algebra of the scale factor
$p$ with holonomies along the symmetry directions fails to be almost periodic functions, as
well as holonomies. This means that one cannot construct the
quantum algebra in the same Hilbert space (Bohr compactification of a real line).  
One of the possibilities is to abandon the $A$ variable and use the connection $\gamma K^i_a$ (which for the $k=-1$ is diagonal) \cite{17}. Then holonomies along left invariant vector fields $\baza{a}{i} \partial_a$ are in the form
\begin{equation}\label{hol}
h_i^{(\mu)}=\porder \exp \left(\int_0^{\mu} ds \gamma K^k_a (\bazaa^a_i) \tau_k \right) =e^{\mu \tilde{c} \tau_i}=\one \cos\frac{\mu \tilde{c}}{2} + 2\tau_i
\sin\frac{\mu \tilde{c}}{2} ,
\end{equation}
where $\mu$ is the length of an edge.
Now it is easy to build quantum algebra of basic operators. Quantum version of the Classical Possion bracket 
$\{ p, e^{\frac{i \mu c}{2}} \}=-i\mu \frac{8 \pi G}{6}e^{\frac{i \mu c}{2}}$ (after rescaling $\tilde{c}=V_0^{-1/3}c$ and $\tilde{p}=V_0^{-2/3}p$) 
is as follows
\begin{equation}
[\hat{p},\widehat{e^{\frac{i \mu c}{2}}}]=\mu \frac{8\pi G\hbar \gamma}{6} \widehat{e^{\frac{i \mu c}{2}}}.
\end{equation}
These operators act on vectors form the Hilbert space
$\mathcal{H}^{\rm gr}=L^2(\mathcal{R}_{\rm Bohr},d\mu_{\rm
  Bohr})$. Eigenstates of $\hat{p}$ consistute an orthonormal basis
$\ket{\mu'} \bra{\mu}=\delta_{\mu' \mu}$ in $\mathcal{H}^{\rm gr}$
\be
\hat{p}\bra{\mu} = \mu \frac{8 \pi l_{\rm Pl}^2 \gamma}{6} \bra{\mu},
\ee
where we denoted $G\hbar:=l_{\rm Pl}^2$. The spectrum of $\hat{p}$ is then discrete.
Each state from $\mathcal{H}^{\rm gr}$ can be decomposed in the
$\bra{\mu}$ basis as a countable sum $\bra{\psi}=\sum_{\mu} \psi(\mu) \bra{\mu}$.
The norm of $\bra{\psi}$ is then defined as
\begin{equation}
\ket{\psi}\bra{\psi}=\sum_{\mu} \bar{\psi}(\mu) \psi(\mu).
\end{equation}
Using classical relation between the physical volume of the elementary cell
and the scale factor $V=|p|^{3/2}$ one can easily construct the volume operator
which is also diagonal in the $\bra{\mu}$ basis
\begin{equation}\label{vol}
\hat{V} \bra{\mu}= |\mu|^{3/2} \left(\frac{8 \pi
    \gamma}{6}\right)^{3/2} l_{\rm Pl}^3 \bra{\mu}.
\end{equation}
The holonomy matrix element operator (\ref{hol}) acts as translations in $\bra{\mu}$ basis
\begin{equation}
\widehat{e^{i\frac{\mu' c}{2}}}\bra{\mu}=\bra{\mu'+\mu}.
\end{equation}
%Note, that because of $\widehat{\exp(i\frac{\mu' c}{2})}$ is not weakly continuous in $\bra{\mu}$ there is no operator corresponding to $c$.
\subsection{The Loop --- preparation}
The formula for the scalar constraint (\ref{ham}) is simplified for symmetry
reduced $k=-1$ model to 
\begin{equation}\label{redu-sc}
H_{gr}=-\frac{1}{16\pi G \gamma^2}\int d^3x N(t) \frac{E^a_i E^b_j}{\sqrt{|{\rm
      det}E|}} {\varepsilon^{ij}}_k (\Lambda^k_{ab} -\gamma^2 \Omega^k_{ab})  ,
\end{equation}
where the part of the curvature proportional to the $\gamma^2$ does not have any dynamical degrees of freedom ($\Omega^k_{ab}=2\partial_{[a} \Gamma^k_{b]} +
{\varepsilon_{ij}}^k \Gamma^i_a \Gamma^j_b$, where $\Gamma$ is defined
in (\ref{reduced})). The curvature 2-form of the $A-\Gamma=\gamma K$ connection reads
\begin{equation}\label{curvat}
\Lambda^k_{ab}=\partial_{[a} \gamma K^k_{b]} + {\varepsilon_{ij}}^k \gamma^2
K^i_a K^j_b=
(-{C^k}_{ij}\tilde{c} + \tilde{c}^2{\varepsilon_{ij}}^k)\ \bazad{i}{a}
\ \bazad{j}{b} ,
\end{equation}
where $C^k_{ij}$ are defined in (\ref{structure}). Naively 
we could introduce the loop such that its holonomy gives two
components, one proportional to structure constants ${C^k}_{ij}$ of
symmetry algebra and the second proportional to structure constants of
$su(2)$ algebra ${\varepsilon_{ij}}^k $. However, putting equation
(\ref{curvat}) into scalar constraint (\ref{redu-sc}) we find 
\begin{equation}\label{dropout}
E^{a}_i E_j^{b} {\varepsilon^{ij}}_k \Lambda^k_{ab} = E^{a}_i
E_j^{b}{\varepsilon^{ij}}_k {\varepsilon_{lm}}^k \gamma^2 K^l_a K^m_b .
\end{equation}
The term in curvature $\Lambda$ proportional to ${C^k}_{ij}$
vanishes. The only term proportional to $\tilde{c}^2 {\varepsilon_{ij}}^k $
contributes to classical Hamiltonian which generates
dynamics. Let us denote 
\begin{equation}\label{cureff}
{\Lambda_{\rm eff}}^k_{ab}={\varepsilon_{ij}}^k \tilde{c}^2 \
\bazad{i}{a} \ \bazad{j}{b} .
\end{equation}
Notice that in (\ref{curvat}) there are only three possibilities for each
values of indecies $i,j$ and $k$: ${C^k}_{ij}=0$ and
${\varepsilon_{ij}}^k \neq 0$, ${C^k}_{ij}\neq 0$ and
${\varepsilon_{ij}}^k = 0$ or ${C^k}_{ij}=0$ and ${\varepsilon_{ij}}^k
= 0$. There are no such $i,j,k$ for which ${C^k}_{ij}$ and
${\varepsilon_{ij}}^k$ contribute at the same ``time''. This is very
different situation to $k=+1$ model, where the terms proportional
to $\tilde{c}$ and $\tilde{c}^2$ contribute to the one and the same component
of the curvature 2-form. Moreover, from (\ref{dropout}) it is clear that the
term proportional to ${C^k}_{ij}$ drops out in the scalar
constraint. It is enough when we find the loop corresponding
only to the $\tilde{c}^2$ term in the curvature (\ref{curvat}), namely
to the (\ref{cureff}) .

\subsection{The Loop}
Let us now construct such a loop. We use technics developed in
\cite{19,18}. The idea is to use the fact that left invariant
vector commute with the right invariant ones. The left inv. fields are defined
in (\ref{exact}) and the right inv. vector fields have the form \cite{27}
\begin{equation}\label{right}
\rig_1=\partial_{x^1}-x^2\partial_{x^2}-x^3\partial_{x^3}, \quad
\rig_2=\partial_{x^2}, \quad \rig_3=\partial_{x^3} .
\end{equation}
It is easy to show, that $[\bazaa_i,\rig_j]=0$ for every $i$ and $j$.
From the geometric interpretation of a Lie bracket of two vector fields
it is clear that arbitrary pair of left and right invariant vector fields
define a closed curve. Moreover, integral curves of those fields
define in a natural way a surface spanned on the loop. In order to define
coordinates on this surface we use a well known fact that every
vector field on a given manifold generates one-parameter group of
diffeomorphisms $\phi(t)$ which maps given point on a manifold
$\vec{x}_0$ to $\vec{x}(t)$ 
\begin{equation}
\phi^{(t)} \vec{x}_0 = \vec{x}(t) .
\end{equation}
If given vector field has a form $V=f^a(x)\partial_a$ (where $f^a(x)$ are
components in given coordinates) such a one-parameter group can be derived from the following condition
\begin{equation}
f^a(x) = \frac{dx^a}{dt} .
\end{equation}
Lets us consider now an arbitrary point $\vec{x}_0=(x^1_0,x^2_0,x^3_0)$ on
$\Sigma$ in the coordinate chart given by (\ref{exact}) and (\ref{right}).
One-parameter diffeomorphism generated by vector fields $\bazaa_2$ and
$\rig_3$ can be written as
\begin{align}\label{flows}
\phi_{(\bazaa_2)}^{(t)}(\vec{x}_0)&=(x^1_0,te^{-x^1_0}+ x^2_0,x^3_0) \nonumber
\\ \phi_{(\rig_3)}^{(s)}(\vec{x}_0)&=(x^1_0,x^2_0,s+ x^3_0) .
\end{align}
The holonomy (with respect to $\gamma K$) along left invariant vector
fields $\bazaa_i$ is simple to calculate (\ref{hol}). What about the
holonomy along right invariant fields? If we start from some point
$\vec{x}_0$ on $\M$, using the formula (\ref{exact}) and (\ref{right}) we will find that the
holonomy along $\rig_3$ has a form
\begin{equation}\label{holrig}
h_{(\rig_3)}=\exp(s e^{x^1_0} \ \tilde{c} \tau_3) .
\end{equation}
Such a holonomy depends on a starting point $x^1_0$ (notice, that the
length of integral curve of $\rig_3$ with respect to the background metric $l_{(\rig_3)}=\int
  \sqrt{\q_{ab}\rig_3^a \rig_3^b}=se^{x^1_0}$).
Now, the loop is defined as follows: We start the holonomy around the loop from an arbitrary point $\vec{x}_0$ on
  $\M$. Using \ref{flows} we get\\
\begin{enumerate}
\item From $(x_0^1,x_0^2,x_0^3)$ we move along $\bazaa_2$ to the point
$(x_0^1, te^{-x^1_0}+x^2_0, x^3_0)$ 
\item From $(x_0^1, te^{-x^1_0}+x^2_0,x^3_0)$ we move along $\rig_3$ to
the point $(x_0^1, te^{-x^1_0}+x^2_0,s+x^3_0)$ 
\item From $(x_0^1, te^{-x^1_0}+x^2_0,s+x^3_0)$ we move to the point
  $(x_0^1,x^2_0,s+x^3_0)$ along $\bazaa_2$ but in the opposite direction
  than in 1) 
\item From $(x_0^1,x^2_0,s+x^3_0)$ we move to the starting point
  $(x_0^1,x^2_0,x^3_0)$ along $\rig_3$, but in opposite direction than in
  2). 
\end{enumerate}
What about the area of the surface spanned by $\bazaa_2$ and $\rig_3$?
  The determinant of a metric tensor pulled back to the surface depends on the point
  of $\M$
\begin{equation}
h:={\rm det} (h_{ab})=e^{2x^1_0} 
\end{equation}
and the area (with respect to the background metric) is
\begin{equation}\label{ar}
{\rm Ar}=\int dt \int ds \cdot \sqrt{h} = ts e^{x^1_0} .
\end{equation}
If we take the length of an integral curve generated by $\bazaa_2$ to be equal
to $\mu$, we can always choose such $s$ in (\ref{ar}) and
(\ref{holrig}) that $s e^{x^1_0}=\mu$. Physical area of the
surface can be constrained to be minimal $Ar_{\rm phy}=\tilde{p}
\bar{\mu}^2=\Delta$ (see \cite{15,18} for details). 
Keeping this in mind and using (\ref{holrig}), (\ref{hol}) and $\bar{\mu}$
condition we get a holonomy around the loop
\begin{equation}\label{loop}
h_{23}^{(\bar{\mu})}=e^{- \bar{\mu} c \tau_3} e^{-\bar{\mu} c \tau_2} e^{\bar{\mu}
  c \tau_3} e^{\bar{\mu} c \tau_2} .
\end{equation}
As in \cite{15,18} shrinking the loop to zero we get the curvature 2-form
\begin{equation}
\bazaa^a_2 \bazaa^b_3 {\Lambda_{\rm eff}}^k_{ab}=-2 \lim_{\bar{\mu} \to 0}{\rm Tr} \frac{\tau_k h_{23}
}{V_0^{2/3}\bar{\mu}^2}=\lim_{\bar{\mu} \to 0} \frac{\sin^2(\bar{\mu} c)}{V_0^{2/3}\bar{\mu}^2} \delta^k_1 .
\end{equation}
Because of homogeneity and isotropy $\bazaa^a_2 \bazaa^b_3 {\Lambda_{\rm eff}}^k_{ab}$ determines the
 ${\Lambda_{\rm eff}}^k_{ab}$ completely 
\begin{equation}\label{regul}
{\Lambda_{\rm eff}}^k_{ab} = \lim_{\bar{\mu} \to 0} \frac{\sin^2 \bar{\mu} c}{V_0^{2/3}\bar{\mu}^2}
{\varepsilon_{ij}}^k \ \bazad{i}{a} \ \bazad{j}{b} .
\end{equation}
Notice that when we shrink our loop to a point $\bar{\mu} \to 0$ we
recover the important part of curvature 2-form (\ref{cureff}) and this
is all we need. Since $\sin\bar{\mu}c$, as well as $\bar{\mu}^{-2}$, is a well defined operator in kinematical Hilbert space, the curvature(\ref{regul}) corresponds to well defined operator in $\mathcal{H}^{\rm gr}=L^2(\mathcal{R}_{\rm Bohr},d\mu_{\rm  Bohr})$.

\subsection{Quantum Dynamics}
Using results from the previous section we can write classical scalar constraint regularized and rescaled by a factor of $16 \pi G$
\begin{equation}\label{regu}
C_{\rm gr}^{\rm reg}=-\frac{6}{\gamma^2} \sqrt{|p|}
(\frac{\sin^2 \bar{\mu}c}{\bar{\mu}^2} - V_0^{2/3}\gamma^2),
\end{equation}
where we have used the rescaled $c$ and $p$ variables. While the term 
\begin{equation}\label{sin}
\sin\bar{\mu}c=\frac{1}{2i}(\exp(i\bar{\mu}c) - \exp(-i\bar{\mu} c))
\end{equation}
corresponds to the well defined operator in $\mathcal{H}^{\rm gr}=L^2(\mathcal{R}_{\rm Bohr},d\mu_{\rm Bohr})$ 
(i.e. translations in volume $\nu=K{\rm sgn}(\mu)|\mu|^{3/2}$, see \cite{15,18} for details) the $\sqrt{p}$ is 
quantized from the classical expression in the spirit of the full LQG in the following manner
\begin{equation}\label{sqrtp}
{\rm sgn}(p)\sqrt{|p|}\ =\ \frac{4}{3\kappa \gamma \bar{\mu}}
\sum_k \mathrm{Tr} \left( h_k^{(\bar{\mu})} \{h_k^{(\bar{\mu})-1},V \}\tau_k\right).
\end{equation}
When we put equations (\ref{regu}), (\ref{sin}) and (\ref{sqrtp}) together we get $\hat{C}_{\rm gr}$ operator. Its action on state $\bra{\psi}=\sum_{\nu}\psi(\nu)\bra{\nu}$ is given by
\begin{equation}\label{quantC}
\hat{C}_{\rm gr}\psi(\nu)=f_{(+)}(\nu) \psi(\nu + 4) + f_{(0)}(\nu)\psi(\nu) + f_{(-)}(\nu)\psi(\nu - 4),
\end{equation}
where the functions $f_{(\pm)}$ are defined as
\begin{align}
f_{(+)}(\nu)&=\frac{27}{16}\sqrt{\frac{8\pi}{6\gamma^3}}K l_{\rm Pl} |\nu+2| | |\nu+1| - |\nu+3| |, \nonumber \\
f_{(-)}(\nu)&=f_{(+)}(\nu-4),  \\
f_{(0)}(\nu)&=-f_{(+)}(\nu)-f_{(-)}(\nu) + A(\nu) \nonumber
\end{align}
and $A(\nu)$ is
\begin{equation}
A(\nu)=3 V_0^{2/3} \sqrt{\frac{8 \pi \gamma}{6}} l_{\rm Pl} \left(\frac{|\nu|}{K}\right)^{1/3} ||\nu+1| - |\nu-1||.
\end{equation}
This way we have found the same scalar constraint operator as the one in \cite{17}!
It is then possible to interpret the Vandersloot \cite{17} operator in the spirit
of the full (LQG) theory: the curvature 2-form is replaced by the holonomy
along a closed curve in the crucial scalar constraint operator. However, the holonomy used in the present paper and in \cite{17} is considered as a function of $\gamma K$ rather then $A$ variable.

\section{Properties of the quantum scalar constraint operator -- Universe with negative cosmological constant}
The \ref{quantC} operator defined in the previous section has the following properties
\begin{itemize}
\item It is densely defined in $\mathcal{H}^{\rm gr}=L^2(\mathcal{R}_{\rm Bohr},d\mu_{\rm Bohr})$ with the domain
\begin{equation}
\Do = \left\{ \bra{\psi} \in \mathcal{H}^{\rm gr}: \bra{\psi}=\sum_{i=1}^n a_i \bra{\nu} , \ a_i \in \C, n \in \Na \right\} ,
\end{equation}
where $\bra{\nu}$ is volume eigenstate.
\item The operator $\hat{C}_{\rm gr}$ preserves every subspace $\mathcal{H}_{\epsilon}$ of $\mathcal{H}^{\rm gr}$
\be 
\H_{\epsilon}={\rm Span} \ \bra{\epsilon + 4n} \in \mathcal{H}^{\rm gr}, n \in \Na
\ee
where $\epsilon$ is an arbitrary real number. We have then the following decomposition into orthogonal subspaces
\be
\mathcal{H}^{\rm gr}=\overline{\bigoplus_{\epsilon} \H_{\epsilon}}.
\ee
\item  $\hat{C}_{\rm gr}$ is symmetric with respect to scalar product
\be
\ket{\psi} \bra{\phi}=\sum_{\nu} \bar{\psi}(\nu) \phi(\nu).
\ee
\end{itemize}
\subsection{Negative cosmological constant}
Classical expression for the cosmological constant has a form $C_{\Lambda}= 2 {\rm sgn}(\Lambda)|p|^{3/2} |\Lambda|$ (do not confuse $\Lambda$ 
with curvature in (\ref{redu-sc})) and its contribution to scalar constraint is of the following form
\be
{C'}_{\rm gr}=-\frac{6}{\gamma^2} \sqrt{|p|} (c^2 - V_0^{2/3}\gamma^2)+  2 {\rm sgn}(\Lambda) |p|^{3/2} |\Lambda| .
\ee
Because the volume operator $\hat{V}=\hat{|p|}^{3/2}$ (\ref{vol}) is known, it is simple to define $\hat{C}'_{\rm gr}$ operator
\begin{equation}\label{sclambda}
\hat{C}'_{\rm gr}\psi(\nu) = \hat{C}_{\rm gr}\psi(\nu) + 2 {\rm sgn}(\Lambda)|\Lambda| \left(\frac{8\pi \gamma}{6} \right)^{3/2} l_{\rm Pl}^3 \frac{|\nu|}{K} \psi(\nu),
\end{equation}
where we used the spectrum of the volume operator in terms of $\nu$ (see \cite{15} for details). Let us now fix ${\rm sgn}(\Lambda)=-1$.
For the negative cosmological constant the following theorem holds: \\
{\bf Theorem} : The operator $\hat{C}'_{\rm gr}$ defined in the domain $\Do$ is essentially self-adjoint.\\
{\bf Proof} : If we rewrite the $\hat{C}'_{\rm gr}$ in the following form
\be
\hat{C}'_{\rm gr}=\underline{\hat{C}} + \hat{C}_0,
\ee
where $\hat{C}_0$ is essentially self-adjoint, then in order to prove the
theorem it is enough to show that
\be\label{proof}
\|\underline{\hat{C}}\psi \|^2 \leq  \| \hat{C}_0 \psi\|^2 + \beta 
\|\psi \|^2
\ee
for each $\psi \in \Do$ and some constant $\beta$ (\cite{20} V.4.6). The action of (\ref{sclambda}) can be written as
\begin{equation}
\hat{C}'_{\rm gr}\psi(\nu) = \underline{\hat{C}}\psi(\nu) + \hat{C}_0 \psi(\nu),
\end{equation}
where
\begin{align}\label{comp}
\underline{\hat{C}}\psi(\nu)&=f_{(+)}(\nu)\psi(\nu+4)+f_{(-)}(\nu)\psi(\nu-4) \nonumber \\
\hat{C}_0 \psi(\nu)&=\left(-f_{(-)}(\nu)-f_{(+)}(\nu) + A(\nu) - 2|\Lambda| \left(\frac{8\pi \gamma}{6} \right)^{3/2} l_{\rm Pl}^3 \frac{|\nu|}{K}\right)\psi(\nu).
\end{align}
$\hat{C}_0$ is multiplication operator so it is obviously essentially self-adjoint. For the norm of $\underline{\hat{C}}$ operator the following inequality holds
\begin{equation}\label{ineq}
\|\underline{\hat{C}}\psi \|^2 = \| (f_{(+)}U_4+ f_{(-)}U_{-4} )\psi
\|^2 \leq 2 \ket{\psi} (f_{(+)}^2 + f_{(-)}^2) \bra{\psi} ,
\end{equation}
where $U_{\pm 4}$ is a unitary translation operator in $\nu$ representation defined by $\exp(\pm 2i\bar{\mu}c)$ (see \cite{15,18} for details).
The (\ref{ineq}) was derived form the inequality $\| u+w \|^2 \leq 2\| u \|^2 + 2 \| w \|^2 $. To conclude, the condition (\ref{proof}) is enough to show that $C_0^2$ (from \ref{comp}) can be written as follows
\be
C_0^2=2f_{(+)}^2 + 2f_{(-)}^2 + f_1 + f_0 ,
\ee
where $f_1>0$ is a function coming from square of (\ref{comp}) and $f_0>0$ is a bounded function which we can always add and it does not change the self-adjointness of $\hat{C}'_{\rm gr}$.

\section{Conclusions}
In this paper we have found a nice analogue of square used in
\cite{15}. Because of the non-commuting character of left invariant fields
$\bazaa_i$ in the hyperbolic $k=-1$ geometry, the loop was constructed
using both left and right invariant fields as in \cite{19,18}. The
important feature of this loop is very natural implementation of 
so--called $\bar{\mu}$ condition (i.e. the physical area of the loop
is constrained to be minimal and equal to the quantum of area
\cite{11} which leads to improved dynamics \cite{15}). Perhaps it
seems surprising that our quantum loop leads to exactly the same operator as introduced by Vandersloot in \cite{17}. This comes form the
fact that the trace of holonomy around our closed curve (\ref{loop})
is precisely the same as the trace of holonomy around the curve generated by each pair
$\bazaa_i, \bazaa_j$ for $i \neq j$ which is not closed as was pointed in \cite{17} (page 8).
Because our two scalar constraint operators are exactly the same, the
correct semi-classical limit of the quantum theory numerically
established in \cite{17} is completely insensitive with respect to our
results. Moreover, from the point of view of quantum theory there are no differences
between Vandersloot model and ours. There are differences in initial
concepts, but they lead to the same quantum theory. However,
assumptions presented in this paper are more natural from the full theory
point of view.

%Important feature of our and Vandersloot model is using $\gamma K$
%holonomies rather then $A=\Gamma + \gamma K$. The main difficulty
%comes from the non-diagonal character of $A$ connection.
In section 4 we have found essentially self-adjoint operator
corresponding to scalar constraint with negative cosmological
constant, but what is the situation when $\Lambda=0$? What about more physical case of the positive cosmological constant?
Unfortunately the theorem described in (\cite{20} V.4.6) can not be
applied to that case due to the fact that inequality (\ref{proof}) no
longer holds for $\Lambda \ge 0$. Moreover, the similar problem arises in the $k=0$ and $k=1$ models with positive cosmological constant when one wants to apply the above theorem. We hope that future investigations give answer to the question about self-adjoint extensions of scalar constraint operators with $\Lambda > 0$.

%Moreover, broken inequality (\ref{proof})
%suggest that operator (\ref{sclambda}) in the $\Lambda \ge 0$ case would not be essentially self-adjoint,
%but the exact proof is not known to the present author. 
% One can say that all three FRW models have the same kinematical (and in some cases also physical) Hilbert space and every scalar constraint operator has a geometrical orgin.

\section{Acknowledgments}
I would like to thank Abhay Ashtekar for suggestions and drawing my attention to $k=-1$ model. I would also like to thank Jerzy Lewandowski, Kevin Vandersloot and especially Wojciech Kami\'nski for important discussions.


\begin{thebibliography}{99}
\bibitem{12} Ashtekar A, Bojowald M and Lewandowski J 2003, Mathematical structure of loop quantum cosmology, \emph{Adv. Theo. Math. Phys.} \textbf{7}, 233-268 (\emph{Preprint} gr-qc/0304074)
\bibitem{3} Ashtekar A and Lewandowski J 2004, Background independent quantum gravity: A status report, \emph{Class. Quant. Grav.} \textbf{21}, R53-R152  (\emph{Preprint} gr-qc/0404018)
\bibitem{11} Ashtekar A and Lewandowski J 1997, Quantum theory of geometry: I. Area operators, \emph{Class. Quantum Grav.} \textbf{14} %No 1A (1997)
A55-A81
\bibitem{13} Ashtekar A, Paw{\l}owski T and Singh P 2006, Quantum nature of the bing bang: An analytical and numerical investigation, \emph{Phys. Rev.} D\textbf{73} 124038, (\emph{Preprint} gr-qc/0604013)
\bibitem{14} Ashtekar A, Paw{\l}owski T, and Singh P 2006, Quantum nature of the big bang, \emph{Phys. Rev. Lett.} \textbf{96}, 141301 (\emph{Preprint} gr-qc/0602086)
\bibitem{15} Ashtekar A, Paw{\l}owski T and Singh P 2006, Quantum Nature
of Big Bang: Improved dynamics, \emph{Phys. Rev.} D \textbf{74} 084003 ,\emph{Preprint} gr-qc/0607039
\bibitem{19} Ashtekar A, Paw{\l}owski T, Vandersloot K and Singh P 2007,
Loop quantum cosmology of $k=1$ FRW models, \emph{Phys. Rev.} D \textbf{75} 024035, \emph{Preprint} gr-qc/0612104
\bibitem{4} Bojowald M 2000, Loop quantum cosmology: I. Kinematics, \emph{Class. Quantum Grav.} \textbf{17} %No 6 (21 March 2000)
1489-1508
\bibitem{5} Bojowald M 2000, Loop quantum cosmology: II. Volume operators, \emph{Class. Quantum Grav.} \textbf{17} %No 6 (21 March 2000)
1509-1526
\bibitem{6} Bojowald M 2001, Loop quantum cosmology: III. Wheeler-DeWitt operators, \emph{Class. Quantum Grav.} \textbf{18} %No 6 (21 March 2001)
1055-1069
\bibitem{10} Bojowald M 2001, Absence of singularity in loop
quantum cosmology, \emph{Phys. Rev. Lett.} \textbf{86}, 5227-5230
\bibitem{9} Bojowald M 2002, Isotropic loop quantum cosmology, \emph{Class. Quantum Grav.} \textbf{19} %No 10 (2002)
2717-2741
\bibitem{26} Bojowald M 2005, Loop quantum cosmology, \emph{Liv. Rev. Rel.} {\bf 8}, 11 (\emph{Preprint} gr-qc/0601085)
\bibitem{16} Bojowald M and Vandersloot K 2003, Loop Quantum Cosmology, Boundary Proposals and Inflation, \emph{Phys.Rev.} D \textbf{67} 124023
\bibitem{18} Szulc {\L}, Kami\'nski W and Lewandowski J 2007, Closed FRW
  model in Loop Quantum Cosmology, \emph{Class. Quantum Grav.} \textbf{24} 2621-2635
\bibitem{20} Kato T 1980, Perturbation Theory for Linear Operators, (Springer-Verlag)
%\bibitem{7} Kiefer C 2006, Quantum Gravity, (Oxford: Claredon Press, International Series of Monographs on Physics ; 124)
%\bibitem{8} Kodama H 1992, Quantum Gravity by the Complex Canonical Formulation, \emph{Int. J. Mod. Phys.} D1 439-524 (\emph{Preprint} gr-qc/9211022)
%\bibitem{JD} Reed M, Simon B 1978, Methods of Modern Mathematical Physics Vol. IV (New York: Academic Press)
\bibitem{2} Rovelli C 2004, Quantum Gravity, (Cambridge: Cambridge University Press)
\bibitem{1} Thiemann T, Introduction to Modern Canonical Quantum General Relativity (Cambridge: Cambridge University Press)\bibitem{17} Vandersloot K 2007, Loop quantum cosmology and the k = -
  1 RW model, \emph{Phys. Rev.} D \textbf{75} 023523, (\emph{Preprint} gr-qc/0612070)
\bibitem{27} Demianski M 1979, Physics of the Expanding Universe, Lecture Notes in Physics (109) (Springer-Verlag, Berlin Heidelber, New York)
%2nd edition
%\bibitem{24} Ashtekar A, Fairhurst S and Willis J L, Quantum gravity, shadow states and quantum mechanics, Class. Quantum Grav. 20 No 6 (2003) 1031-1061, gr-qc/0207106
%\bibitem{25} Ashtekar A and Bojowald M, Quantum geometry and the Schwarzschild singularity, Class. Quantum Grav. 23 No 2 (2006) 391-411
\end{thebibliography}
\end{document}